# Linking Received Packet to the Transmitter Through Physical-Fingerprinting of Controller Area Network

Omid Avatefipour
Department of Electrical and Computer Engineering
University of Michigan-Dearborn
Dearborn, United States
oavatefi@umich.edu

Azeem Hafeez
Department of Electrical and Computer Engineering
University of Michigan-Dearborn
Dearborn, United States
azeemh@umich.edu

Muhammad Tayyab
Department of Electrical and Computer Engineering
University of Michigan-Dearborn
Dearborn, United States
tayyab@umich.edu

Hafiz Malik
Department of Electrical and Computer Engineering
University of Michigan-Dearborn
Dearborn, United States
hafiz@umich.edu

*Abstract*— The Controller Area Network (CAN) bus serves as a legacy protocol for in-vehicle data communication. Simplicity, robustness, and suitability for real-time systems are the salient features of the CAN bus protocol. However, it lacks the basic security features such as massage authentication, which makes it vulnerable to the spoofing attacks. In a CAN network, linking CAN packet to the sender node is a challenging task. This paper aims to address this issue by developing a framework to link each CAN packet to its source. Physical signal attributes of the received packet consisting of channel and node (or device) which contains specific unique artifacts are considered to achieve this goal. Material and design imperfections in the physical channel and digital device, which are the main contributing factors behind the device-channel specific unique artifacts, are leveraged to link the received electrical signal to the transmitter. Generally, the inimitable patterns of signals from each ECUs exist over the course of time that can manifest the stability of the proposed method. Uniqueness of the channel-device specific attributes are also investigated for time- and frequency-domain. Feature vector is made up of both time and frequency domain physical attributes and then employed to train a neural network-based classifier. Performance of the proposed fingerprinting method is evaluated by using a dataset collected from 16 different channels and four identical ECUs transmitting same message. Experimental results indicate that the proposed method achieves correct detection rates of 95.2% and 98.3% for channel and ECU classification, respectively.

## I. INTRODUCTION

The Controller Area Network (CAN) bus protocol is widely used for embedded systems networking. It finds a wide range of applications from automotive, aerospace, agriculture, medical devices, and even in some of the home and commercial appliances [1].

A modern vehicle contains many different computing devices, known as Electronic Control Unit (ECU), which are responsible for sensing and controlling actuators [2]. Virtually, all functionalities in the modern automobiles ranging from engine control to braking, lighting, driver safety, antilock brake systems (ABS) and the parking assist systems are achieved through these ECUs [3]. These ECUs communicate with each other through different networks. If the communication on these networks is not secured, it can pose a serious threat to the safety of the passengers.

The CAN-bus has been a de-facto standard for communication as an in-vehicle network for over 30 years. By design, the CAN-bus lacks basic security features such as message authentication option which makes it vulnerable to a variety of spoofing attacks [4]. For example, in the absence of effective message authentication, a single compromised ECU allows the attacker to take full control of the vehicle by injecting spoofed messages [2,5,6]. Since CAN packets contain no authenticator field, any ECU on the network can impersonate the other ECUs in the network. This provides a broad range of internal as well as external attack surfaces [7]. An adversary can leverage the CAN-Bus protocol vulnerabilities to launch various attacks leading to malfunctioning of the vehicle. Data encryption-based solutions are proven to be inefficient for the CAN-Bus protocol [7]. Lack of the channel encryption provides the adversary an opportunity to sniff the network traffic by simply plugging in a low-price hardware leading to the replay attacks [8].

Attack surfaces are growing by the course of time which gives rise to develop the effective protection of CAN-bus communication from malicious attackers as a challenging task. The automakers are aiming for a fully-connected intelligent vehicle which makes secure in-vehicle communication problem even more complicated. Recently, researchers have proposed many solutions for in-vehicle networks security at different layers e.g. physical layer [14,23] and data link layer by using various types of message authentication methods [7,15,16,22].

In this paper, we propose a method to link the received packet to its transmitter based on the unique physical properties of the signal. The proposed physical-fingerprinting-based method exploit unique artifacts both at the digital device (ECU) level and in the physical channel (e.g., CAN-bus). Material and design imperfections in the channel and the transmitter are the main contributing factors behind these unique artifacts. The physical channel unique artifacts, which are used to link received electrical signal to the source (or transmitting) ECU, are considered in this study. More specifically, the proposed method exploits physical channel dependent attributes for linking received signals (message) to the transmitting device. The proposed method can be leveraged as an identification method in such a way that if an adversary tries to send a malicious message either from an external ECU or by changing the cables, it can be distinguished as a malicious activity and based on the defined safety specifications proper actions can be performed. Even if an adversary uses the legitimate message identifier (e.g. shut down engine), since he/she is sending that message from an external ECU, the proposed method can detect that signal has not originated from the legitimate one because the signal will not pair with the ECU that should have generated that message. It has been observed that uniqueness of the physical attributes exists both in time and frequency domain. In this paper, a feature vector consisting of 11 time and frequency domain statistical signal attributes including higher-order moments, spectral flatness measure, minimum, maximum, and irregularity K are considered to capture the channel and the transmitter dependent uniqueness. A multi-layer neural network based classifier is trained and tested for source ECU and the source channel. Experimental results indicate that the proposed attributes can be used to classify different channels and ECUs. Performance of the proposed fingerprinting method is evaluated on a dataset collected from 16 different channels and four identical ECUs transmitting the same message. Experimental results demonstrate that the proposed method achieves correct detection rates of 95.2% and 98.3% for channel and ECU classification, respectively.

The rest of this paper is organized as follows: Section II presents an overview of CAN-bus protocol. Section III provides a brief overview of the related work in the area of CAN-Bus security and authentication techniques. Details of the proposed method are outlined in Section IV and experimental results and analysis are explained in Section V then conclusion and future directions are discussed in Section VI.

## II. CAN-BUS PROTOCOL: AN OVERVIEW

The CAN-Bus is the broadcasting based communication topology where each node receives the transmitted messages. However, each node accepts the messages with a particular ID and discards the others. Depending on the node configuration and its functionalities, the communication of the network is filtered which means that each node only accepts particular message not all the incoming messages. Message transmission on the network is event-driven [11]. The CAN messages are identified based upon the identifier field, denoted as ID. The ID is used for prioritizing the messages as well to avoid the collision in case of contention between nodes to transmit at the same time. The message with the lower ID has higher priority for winning the contention. For example, if two different nodes tend to transmit the messages with the identifier value of 0x12 and 0xF4 at the same time, the message with ID 0x12 is sent first due to the lower value. There are two formats for CAN-bus namely standard format which has 11-bit identifier and extended format with 29-bit identifier [12]. In automotive industry, differential signal voltage is mostly used for the physical layer signaling using two communication wires e.g. CAN-High and CAN-Low [7]. Shown in Figure 1 is the bit transition and signal voltages of CAN bus communication which includes series of dominant and recessive bits. When a recessive bit (logical 1) is transmitting both CAN-High and CAN-low are driven to the 2.5 volts which indicates that the voltage difference is zero during the transmission of recessive bit and when a dominant bit (logical 0) is transmitted, CAN-High goes to 3.5 volts and CAN-Low goes down to the 1.5 that means the voltage difference in the dominant bit is 2 volts [13]. As a result, if two nodes are trying to publish on the bus simultaneously then dominant bit will win the arbitration. Therefore, it can be concluded that: "the lower the value of identifier is, the higher will be the priority to win the arbitration and publish data on the bus".

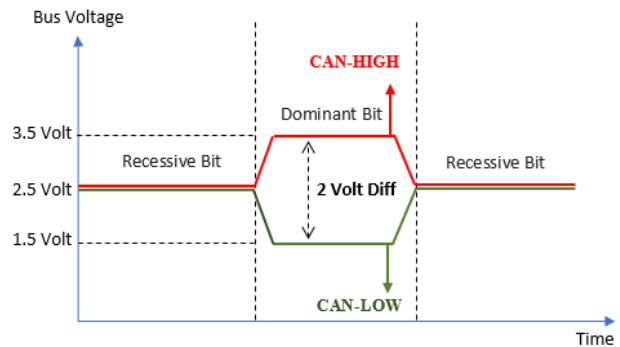

Fig. 1. CAN-bus differential signal representations

## III. RELATED WORK

Recently, the research community has shown a growing interest and attention on CAN-Bus security. For example, Cho and Shin [14] proposed a clock skew based framework for ECU fingerprinting and use it for the development of Clock based Intrusion Detection System (IDS). The proposed clock based fingerprinting method [14] exploited clock characteristic which exists in all digital systems, that is, "*tiny timing error known as clock skew*". The clock skew identification exploits uniqueness of the clock skew and clock offset which is used to identify a given ECU based on clock attributes of the sending ECU. Cho and Shin [14] also developed a prototype of the proposed IDS and demonstrated effectiveness of the proposed CIDS on three different vehicles e.g. Honda Accord, Toyota Camry, and a Dodge Ram.

Wang et al. [7] developed a practical security framework for vehicular systems (VeCure) in which they implemented the message authentication code (MAC) for the CAN-bus. The VeCure method realized by developing a proof-of-concept testbed on Freescale automotive evaluation board. In their method, each node which sends message should also send another 8-byte message for authentication. High computational cost is one of the limitation of the proposed method. For example, it requires 2000 additional clock cycles.

Hiroshi et al. [15] proposed a security authentication monitoring system for CAN-Bus which uses MAC for protecting CAN bus against spoofing attacks. The role of monitoring node in their proposed method is to authenticate each ECU and verified the authentication code which is defined for each CAN message. The modified CAN controller is required to install for their monitoring node to implement the message authentication which transmits an error frame to overwrite spoofed message. Additionally, if the monitoring node is compromised or removed from the bus, the entire network is compromised.

Hazem et al. [16] proposed a Lightweight CAN Authentication Protocol LCAP. The proposed method requires to append a "magic number" which can be generated on the one-way hash function employed in TESLA protocol [21] for the message to be verified from the receiver side. Handshake technique is used for node synchronization and channel security. It requires 2 bytes of the data field for the authentication code which only creates small overhead for message authentication code exchange among the nodes. However, since the LCAP introduces the new IDs in the network configuration, it requires large address space.

## IV. PROPSED METHOD: CHANNEL RESPONSE BASED ECU IDENTIFICATION

The proposed transmitted identification method relies on the fact that each electronic device (e.g. ECU) and channel impulse response of the physical channel (e.g., CAN-Bus) exhibit unique artifacts which can be used for linking received signal to the sending ECU. More specifically, by extracting the distinguishable statistical features of transmitting signals, the source of the coming message is identified.

Let $S_i(t)$ be the output of the $i^{th}$ ECU and $h_j(t)$ be the impulse response of the $j^{th}$ physical channel between $i^{th}$ ECU and the physical fingerprinting (PhyFin) unit. The physical signal at the input of the PhyFin unit, $y_{ij}(t)$, can be expressed as Equation 1 and Figure 2, respectively.

$$y_{ij}(t) = h_j(t) * S_i(t) \quad (1)$$

where, * denotes convolution operator.

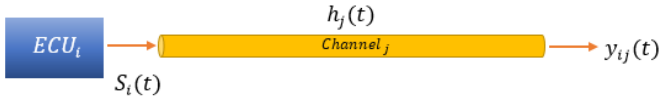

Fig. 2. Physical input signal and channel response

Physical signal at the input of PhyFin unit, $y_{ij}(t)$ is used for linking $y_{ij}(t)$ to its source. Shown in Figure 3 are plots of four waveforms at the output of four different channels when identical message is applied at the input of these channels. It can be observed from Figure 3 that channel impulse response is different for all four channels, which validates our claim of channel specific uniqueness.

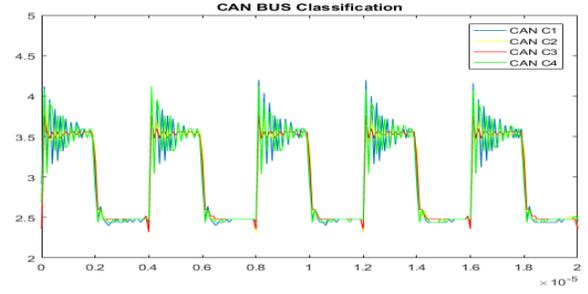

Fig. 3. Waveforms of the received signals from four different CAN-bus channels with identical channel input message.

Various feature extraction methods, both in time and spectral domain are evaluated here. To validate effectiveness of the proposed method here, feature extraction method presented in [17] is considered. To this end, 40-dimensioanl scalar features both in time and spectral domain are extracted using LibXtract - *a library for feature extraction* [18]. The extracted feature set is then analyzed further to select relevant features. FEAST Toolbox is applied [19,] which utilizes the joint mutual information criterion, for ranking the features. Details of the selected time- and frequency-domain features are shown in Table I and Table II, respectively. The feature selection process resulted in an 11-D feature vector for channel and ECU identification.

TABLE I. TIME DOMAIN FEATURE SET

| Feature name | Equation |
|---|---|
| Maximum | $m_{ij} = (Min(y_{ij}(i)) \mid i=1…N)$ |
| Minimum | $M_{ij} = (Max(y_{ij}(i)) \mid i=1…N)$ |
| Mean | $\mu_{ij} = \frac{1}{N} \sum_{i=1}^{N} y_{ij}(i)$ |
| Variance | $\sigma_{ij}^2 = \sqrt{\frac{1}{N-1} \sum_{i=1}^{N} y_{ij}(i) - \mu_{ij}}$ |
| Skewness | $\rho_{ij} = \frac{1}{N} \sum_{i=1}^{N} \left(\frac{y_{ij}(i) - \mu_{ij}}{\sigma_{ij}}\right)^3$ |
| Kurtosis | $\kappa_{ij} = \frac{1}{N} \sum_{i=1}^{N} \left(\frac{y_{ij}(i) - \mu_{ij}}{\sigma_{ij}}\right)^4 - 3$ |

TABLE II. FREQUENCY DOMAIN FEATURE SET

| Feature Name | Equation |
|---|---|
| Spectral Std-Dev | $\sigma_s = \sqrt{(\sum_{i=1}^{N} (y_f(i))^2 * (y_m(i))) / \sum_{i=1}^{N} (y_m(i))}$ |
| Spectral Skewness | $\rho_s = (\sum_{i=1}^{N} y_f(i)(y_m(i)) / \sigma_s^3$ |
| Spectral Kurtosis | $\kappa_s = (\sum_{i=1}^{N} (y_m(i) - C_s)^4 * y_m(i)) / \sigma_s^4 - 3$ |
| Spectrum Centroid | $C_s = (\sum_{i=1}^{N} y_f(i) y_m(i)) / (\sum_{i=1}^{N} y_m(i))$ |
| Irregularity-K | $IK_s = \sum_{i=2}^{N-1} \mid y_m(i) - \frac{y_m(i-1) + y_m(i) + y_m(i+1)}{3} \mid$ |

*$y_m$ and $y_f$ are the magnitude and the frequency vectors respectively*

## A. Experimental Setup

Three different type of channels, GXL, TXL, SAE J1939-15, are used for CAN-Bus. These channels are being used actively in real vehicles. Details of the channel types and channel lengths are outlined as follows:

- GXL primary automotive cable is used for engine compartment where high resistance is required according to SAE J1128. [20]
- TXL is also primary automotive cable used for applications requiring smaller diameters and minimal weight.
- CAN-bus data cables SAE J1939-15 which is used for connecting different ECUs to network.

The technical specification of each channel is provided in Table III. Six (6) channel lengths are considered to realize CAN-Bus with pairs of twisted wires from same manufacturer and gauge. Overall, the experimental setup contains following hardware and software components:

- Four (4) Arduino Uno R2 microcontroller kits
- Four (4) CAN-Bus shield board with MCP2515 CAN-bus controller and MPC2551 CAN transceiver.
- Three (3) different types of Cables (GXL, TXL, and CAN-bus data cable) with multiple lengths: 0.5 meter, 1 meter, 2 meter, 3 meter, 4 meter, and 5 meter.
- Oscilloscope DSO1012A for the voltage samples recording with Sampling Rate of 2GSa/s, 100MHz bandwidth, and 8-bit vertical resolution.
- Script for sending an identical message continuously from different channels and ECUs to observe the unique patterns of signals from each channel and ECU.
- MATLAB R2016a software for statistical data analysis of sampled signals.

## B. Dataset description

Performance of the proposed algorithm is evaluated for both CAN-Bus channel and ECU classification. To this end, physical signal is captured at the output of three different cable families with multiple lengths (0.5 meter, 1 meter, 2 meters, 3 meters, 4 meters, and 5 meters) and four identical ECUs with same input CAN-bus message. To this end, a dataset for the 18 channels and four identical ECUs is collected. For each data collection setting, 144000 (3600*40) samples are collected. For performance evaluation, random partitioning is performed to divide the dataset into the training and test set (Training set: 65%, Test set: 35%). The dataset used here is collected in the same environment i.e. under the same temperature and using an identical message to observe the minute and unique variation of the digital signals.

TABLE III. TECHNICAL SPECIFICATION OF THREE DIFFERENT CABLE FAMILIES

| Type | AWG | Conductor | Insulation | No. of Strands | Temperature | Compliances |
|---|---|---|---|---|---|---|
| GXL | 18 | Bare copper | Cross-linked Polyethylene (XLP) | 16x30 | -40°C -125°C | Ford ESB-(M1L85-A), Chrysler (MS8900), SAE-J-1128. |
| TXL | 18 | Bare copper | Cross-linked Polyethylene (XLP) | 19x30 | -40°C -125°C | Ford (M1L-123A), Chrysler (MS-8288), SAE-J-1560 |
| CAN-bus Data cable | 18 | Bare copper | Cross-Linked Polyolefin (XLPO), Thermoplastic Polyurethane (TPU) | 19x31 | -45°C -125°C | SA J1939-11 Physical Media, RoHS, SAE J1128 performance (fluid, flame propagation) |

## V. EXPERIMENTAL RESULTS AND ANALYSIS

Performance of the proposed method is evaluated through a series of experiments for channel as well as ECU identification. To achieve this goal, a multilayer neural network based classifier is trained on randomly selected 65% data for each channel and ECU. The trained classifier is then employed to test performance of the proposed methods on remaining 35% data. Classification accuracy is used to measure performance of the proposed method.

### A. *Experiment 1: Channel Identification*

The main objective of this experiment is to validate uniqueness of channel specific features. Material and design imperfections for each specific physical channel is the leading factors behind the channel specific unique artifacts. To validate this claim, data is recorded for each cable family and each channel length with identical channel input, transmitted using the same ECU. Specifically, for this experiment *'cable type'* and *'length'* are the only variables. During the training phase, the neural network is trained for classifying three different cable family and six corresponding channel lengths (e.g., GXL: 0.5 meter, GXL: 1 meter, GXL: 2 meter, GXL: 3 meter, GXL: 4 meter, and GXL: 5 meter and so on). A multilayer neural network is trained with "scaled conjugate gradient back propagation" training algorithm, 11 inputs variables (time and frequency domain), 6 outputs which corresponds to different lengths of GXL cable, stopping criteria of Epochs = 2000, gradient = 1e-7, and three hidden layers with 50, 40, and 40 hidden nodes respectively. Shown in Figure 4 is the architecture of the multilayer neural network trained for channel classification.

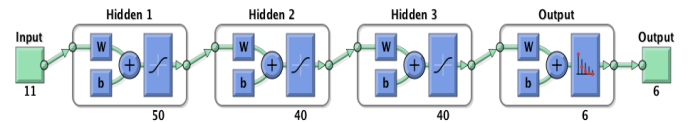

Fig. 4. Neural Network architecture of channel classifier

Shown in Table IV.A and IV.B are the confusion matrices of the channel (C) classification averaged over all cable types for the training and test phase. It can be observed from Table IV that that the proposed method for channel classification achieves overall correct detection rate of 97.6% and 95.2% for the training and test phase, respectively. It can also be noticed that 0.5 meter and 1 meter channels exhibit relatively higher false rates for both training and testing, these false rates can be attributed to the fact that both channel lengths are not very different. The signal characteristics uniqueness exists for each family type cable and the corresponding lengths.

TABLE IV.A. TRAINING CONFUSION MATRIX FOR CHANNEL CLASSIFIER

| | C1 | C2 | C3 | C4 | C5 | C6 | |
|---|---|---|---|---|---|---|---|
| C1 | 365 / 15.6% | 4 / 0.2% | 0 / 0.0% | 0 / 0.0% | 0 / 0.0% | 0 / 0.0% | 98.9% / 1.1% |
| C2 | 30 / 1.3% | 378 / 16.2% | 0 / 0.0% | 0 / 0.0% | 0 / 0.0% | 0 / 0.0% | 92.6% / 7.4% |
| C3 | 2 / 0.1% | 0 / 0.0% | 376 / 16.1% | 12 / 0.5% | 0 / 0.0% | 0 / 0.0% | 96.4% / 3.6% |
| C4 | 1 / 0.0% | 0 / 0.0% | 8 / 0.3% | 382 / 16.3% | 0 / 0.0% | 0 / 0.0% | 97.7% / 2.3% |
| C5 | 0 / 0.0% | 0 / 0.0% | 0 / 0.0% | 0 / 0.0% | 388 / 16.6% | 0 / 0.0% | 100% / 0.0% |
| C6 | 0 / 0.0% | 0 / 0.0% | 0 / 0.0% | 0 / 0.0% | 0 / 0.0% | 394 / 16.8% | 100% / 0.0% |
| | 91.7% / 8.3% | 99.0% / 1.0% | 97.9% / 2.1% | 97.0% / 3.0% | 100% / 0.0% | 100% / 0.0% | **97.6% / 2.4%** |

Target Class (Predicted Class on vertical axis)

TABLE IV.B. TEST CONFUSION MATRIX FOR CHANNEL CLASSIFIER

| | C1 | C2 | C3 | C4 | C5 | C6 | |
|---|---|---|---|---|---|---|---|
| C1 | 176 / 14.0% | 10 / 0.8% | 0 / 0.0% | 0 / 0.0% | 0 / 0.0% | 0 / 0.0% | 94.6% / 5.4% |
| C2 | 22 / 1.7% | 205 / 16.3% | 0 / 0.0% | 0 / 0.0% | 0 / 0.0% | 0 / 0.0% | 90.3% / 9.7% |
| C3 | 3 / 0.2% | 3 / 0.2% | 203 / 16.3% | 9 / 0.7% | 0 / 0.0% | 0 / 0.0% | 93.1% / 6.9% |
| C4 | 1 / 0.1% | 0 / 0.0% | 13 / 1.0% | 197 / 15.6% | 0 / 0.0% | 0 / 0.0% | 93.4% / 6.6% |
| C5 | 0 / 0.0% | 0 / 0.0% | 0 / 0.0% | 0 / 0.0% | 212 / 16.8% | 0 / 0.0% | 100% / 0.0% |
| C6 | 0 / 0.0% | 0 / 0.0% | 0 / 0.0% | 0 / 0.0% | 0 / 0.0% | 206 / 16.3% | 100% / 0.0% |
| | 87.1% / 12.9% | 94.0% / 6.0% | 94.0% / 6.0% | 95.6% / 4.4% | 100% / 0.0% | 100% / 0.0% | **95.2% / 4.8%** |

Target Class

### B. *Experiment 2:* ECU Identification

The purpose of this experiment is to validate that different ECUs even from the same make and model introduce different artifacts while transmitting an identical message. To achieve this goal, dataset for all four ECUs transmitting same messages over the same channel is used. In this experiment, ECU is the only variable while other variables are kept constant. To this end, data for all four ECUs transmitting same messages over the 2-meter unshielded CAN-Bus data cable is used for training and testing. A multilayer neural network classifier is trained with "scaled conjugate gradient back propagation" training algorithm, 11 input variables (both time and frequency domain), 4 outputs which pertains to each ECU, stopping criteria of Epochs = 2000, gradient = 1e-7, and one hidden layer with 20 hidden nodes included. Shown in Figure 5 is the architecture of the multilayer NN trained for channel classification.

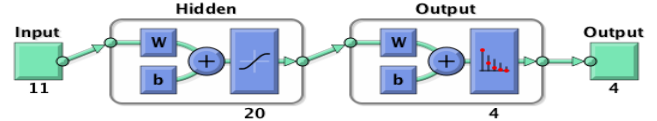

Fig. 5. Neural Network architecture of ECU classifier

Shown in Table V.A and V.B are the classification performance of the proposed system in terms of confusion matrices of the ECU (E) classification for the training and test phases, respectively. It can be observed from Table V that the proposed method for ECU classification achieves overall success detection rate of 99.6% and 98.3% during the training and test phase, respectively.

TABLE V-A. TRAINING CONFUSION MATRIX FOR ECU CLASSIFIER

| | E1 | E2 | E3 | E4 | |
|---|---|---|---|---|---|
| E1 | 389 / 24.9% | 0 / 0.0% | 3 / 0.2% | 0 / 0.0% | 99.2% / 0.8% |
| E2 | 0 / 0.0% | 398 / 25.5% | 0 / 0.0% | 0 / 0.0% | 100% / 0.0% |
| E3 | 3 / 0.2% | 0 / 0.0% | 379 / 24.3% | 0 / 0.0% | 99.2% / 0.8% |
| E4 | 0 / 0.0% | 0 / 0.0% | 0 / 0.0% | 398 / 24.9% | 100% / 0.0% |
| | 99.2% / 0.8% | 100.% / 0.0% | 99.2% / 0.8% | 100% / 0.0% | **99.6% / 0.4%** |

Target Class

TABLE V-B. TEST CONFUSION MATRIX FOR ECU CLASSIFIER

| | E1 | E2 | E3 | E4 | |
|---|---|---|---|---|---|
| E1 | 200 / 23.8% | 0 / 0.0% | 6 / 0.7% | 0 / 0.0% | 97.1% / 2.9% |
| E2 | 0 / 0.0% | 202 / 24.0% | 0 / 0.0% | 0 / 0.0% | 100% / 0.0% |
| E3 | 7 / 0.8% | 0 / 0.0% | 212 / 25.2% | 0 / 0.0% | 96.8% / 3.2% |
| E4 | 1 / 0.1% | 0 / 0.0% | 0 / 0.0% | 212 / 25.2% | 99.5% / 0.5% |
| | 96.2% / 3.8% | 100.% / 0.0% | 97.2% / 2.8% | 100% / 0.0% | **98.3% / 1.7%** |

Target Class

## VI. CONCLUSION

In this study, we have demonstrated that for an identical CAN-Bus message, underlying physical channel leaves inimitable characteristic artifacts in the signals at the channel output. These artifacts are unique to different channel lengths and ECUs. The received physical signal therefore can be used for linking received CAN packet to actual transmitter. Statistical attributes in time and frequency domain are utilized for channel and device identification. The performance of the classification method is evaluated by carrying out the experimental setup for three different CAN-Bus channels with six multiple lengths and four ECUs from the same manufacturer. The experimental results and analysis indicate that the proposed method achieves the satisfactory CAN-Bus channel and ECU identification performance with the overall correction rate of 95.2% and 98.3%, respectively. For the future work, development of an identification platform for security purposes will be investigated to determine whether the received message is from the compromised ECU or legitimate one by leveraging these unique signal characteristics.